\documentclass[runningheads]{svmult}

\usepackage{makeidx}   
\usepackage{graphicx}  
\usepackage{subeqnar}  
\usepackage{multicol}  
\usepackage{physprbb}  
\makeindex             

\begin{document}
\title*{Kinematics of Distant Galaxies from Keck}
\toctitle{Kinematics of Distant Galaxies from Keck}
%
%
\titlerunning{Kinematics of Distant Galaxies}
%
\author{David C. Koo}
%
\authorrunning{David C. Koo}
%
%
\institute{UCO/Lick Observatory and Department of Astronomy and Astrophysics,\\ University of California, \\
Santa Cruz, CA 95064, USA}

\maketitle              

\begin{abstract}
DEEP\index{DEEP}  is a two-phase spectral
survey of faint field galaxies with the Keck Telescopes. The
goals include exploring galaxy formation \index{galaxy formation} 
and evolution \index{galaxy evolution},
mapping distant large scale structures, and
constraining cosmology.
DEEP, since its inception in the early 1990's, has been 
distinguished by an emphasis on studying the kinematics \index{kinematics} 
and masses \index{masses} of 
distant galaxies.
The major DEEP survey  in the second phase (DEEP2) is scheduled to
begin in 2002 and will mainly aim
for a sample of 50,000 galaxies to $I \sim 23$.
Until then, the first
phase of DEEP science programs will have  been concentrating on using 
existing  Keck spectrographs to undertake spectral surveys 
of over 1000 galaxies that have also been observed with HST.
I will highlight the study of rotation curves \index{rotation curves} 
of distant spirals;
the fundamental plane \index{fundamental plane} 
of faint, high-redshift E/S0s\index{ellipticals}
\index{spheroids} \index{early-type galaxies}; 
the narrow velocity widths seen in luminous blue compact galaxies; and 
the diversity of kinematics seen in a small sample of 
high redshift ($z \sim 3$)
galaxies. These DEEP pilot programs have clearly demonstrated the
feasibility, importance, and potential of using kinematics to better
understand distant galaxies. 
\end{abstract}

\section{What is DEEP?}
The first decade of the 21st century promises many new 
surveys of distant galaxies, especially with the advent of a suite of new
8-10~m class, ground-based, optical telescopes.  Besides adding
critical redshifts to data from space and other wavebands, the higher
S/N and spectral-resolutions affordable with 8-10m telescopes
provide three new and quite powerful diagnostics for the analysis of distant
galaxies: internal velocities (i.e., kinematics and
hence dynamical masses when size is also measured); chemical
abundances; and star-formation-rate and stellar-population-age
estimates.  Compared to the traditional parameters of counts, colors,
luminosities, and clustering properties of distant galaxies, these
new diagnostics yield independent probes of galaxy
properties in the early universe and have solid links to theoretical
simulations of galaxy formation.  Moreover, since  both galaxy
evolution and their large scale patterns involve a
complex interplay of diverse galaxy classes,
environments, and physical mechanisms and 
because precision cosmology via the volume test
requires averaging over the fluctuations due to large-scale
clustering, very large samples are essential to
extract reliable results.

To meet the challenge, DEEP\footnote{DEEP: Deep Extragalactic
Evolutionary Probe; more details on participants and programs of DEEP
can be found at URL: {\bf http://www.ucolick.org/ $\sim$deep}} was
initiated over 9 years ago as a spectral survey of 
50,000 \footnote{our original goal of
10,000 has been revised upwards to improve significantly the
reliability of cosmological tests and of large scale structure studies} 
faint field galaxies ,
using the Keck II 10-m Telescope with a new spectrograph DEIMOS
\cite{Dav98} \footnote{ DEIMOS: DEep Imaging Multi-Object Spectrograph; more
information is provided at URL: {\bf http://www.ucolick.org/
$\sim$loen/Deimos/deimos.html}}.
The use of DEIMOS provides a clean division of DEEP into two parts or
phases. The first is a set of pilot-style surveys of relatively small
samples (10's - 1000) of galaxies. These pilot surveys 
exploit the pre-DEIMOS spectrographs available on Keck 
and were designed to determine feasibility and to refine
the scope of DEEP2. DEEP is distinguished by aiming
to gather internal kinematic data in the form of rotation curves or
linewidths, as well as spectral-line measurements sensitive to star
formation rates, gas conditions, stellar-population ages, and
metallicity.

\section{DEEP Highlights on Kinematics}

To maximize the scientific returns for the small samples
from  our phase-one, pilot surveys,
we only observed fields where HST WFPC2 images already exist,
including the HDF and flanking fields \cite{Guz97}, \cite{Low97},
\cite{Phi97}, \cite{Vogt97}; the Groth Strip Survey (GSS) \cite{Koo96}
\cite{Phi02} \cite{Vogt96} \cite{Vogt02}; and Selected Area 68.  Such
HST images provide not only morphology and photometry but also the
structure, size, and inclination data needed to convert kinematic
observations from Keck into direct measures of dynamical mass.

The DEEP data reach to $I \sim 24$ and confirms 
that  DEEP2 is feasible and that kinematics and masses will
be worth the extra effort \cite{Koo00}. 
I will highlight  our studies of distant spirals
via rotation curves; luminous red spheroids and  blue compact galaxies
via velocity dispersions; 
and a few high redshift ($z \sim 3$) ``Lyman-drop'' galaxies for which
spatially-resolved kinematics is possible.

\subsection{Rotation Curves of Distant Spirals}

\begin{figure}[b]
\begin{center}
\includegraphics*[height=2.0in]{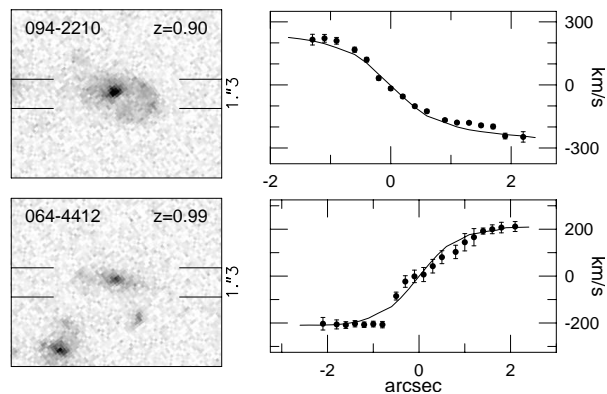}
\end{center}
\caption[]{Examples of the rotation curves measured for two
high redshift galaxies \cite{Vogt97}, 
the upper with total $I \sim 21.4$ and the lower
with $I \sim 22.4$. Besides the ID and redshift at the top, the images
show the orientation and width of the slit.
}
\label{fig1}
\end{figure}

As seen inf Fig.\ref{fig1}, we have clearly demonstrated that
emission-line rotation curves of likely spirals can be measured with
Keck's low resolution spectrograph (LRIS: \cite{Oke95}) to redshifts
near $z \sim 1$ for galaxies as faint as $I \sim 22$ with 1--2 hour
exposures \cite{Vogt97}.  Our new sample of about 100 rotation curves
\cite{Vogt99} \cite{Vogt02} support the original conclusion that the
optical Tully-Fisher relation \index{TF: Tully-Fisher} for spirals
near redshifts $z \sim 1$ show only modest ($< 0.6$mag) changes
relative to that seen locally\cite{Vogt96}, \cite{Vogt97}. These
results appear on the surface to disagree with the claims for more
extensive evolution of 1.5 to 2.0 mag \cite{Rix97} \cite{Sim99}
\cite{Mal99}, but the differences may reflect the selection criteria
adopted. While our high-quality Tully-Fisher sample included galaxies
that were slightly elongated and resolved along the slit, i.e. larger
disk systems, the other samples were generally limited to very blue or
strong emission line targets, some of which may be very compact.  We
are currently expanding the completeness of our kinematic sample by
including emission-line velocity widths in our studies (see
contribution by B. Weiner). More solid conclusions on the Tully-Fisher
relation (velocity vs. luminosity), as well as that of other scaling
relations when one adds size or surface brightness, are critical to
test whether various theories of disk formation are
correct\cite{Fab01}.

%

\subsection{Spheroid/Bulge Evolution}

We have undertaken several  approaches to explore the evolution of
luminous early-type galaxies, two of which exploit kinematics.  
In the first study\cite{Im02}, the luminosity function of over
100 early-type galaxies (E/S0) was derived, where the early-type class
was selected on the basis of B/T being larger than 0.4; low levels of
asymmetries; and high levels of smoothness using the GIM2D software
for structural parameter extractions \cite{Sim02}. Keck redshifts were
complemented by photometric redshifts. The main result is that there
is evidence for roughly 1 magnitude of luminosity brightening back to
redshifts $z \sim 1$, but otherwise no evidence for any dramatic drop
in number density \cite{Im02}. This result supports 
hierarchical models of galaxy formation in an open or
accelerating universe rather than a  high-$\Omega$ one
(e.g., SCDM), in which  early-type galaxies are presumed to have formed 
via merging of spirals in
significant numbers since $z \sim 1$.  In the
second work \cite{Im01}, a more detailed study was made of the blue E/S0s.
Kinematic information, including some very high resolution velocity
widths as measured with an Echelle system (ESI: \cite{She00}), 
were used and we found nearly all to be
low-mass systems, rather than bonafide massive E/S0 galaxies
undergoing intense star formation.

In the third study\cite{Geb01}, we explored the fundamental plane
(luminosity: size: velocity dispersions) of 35 galaxies using LRIS
data and found evidence for nearly 2 mag of luminosity brightening
back to redshifts $z \sim 1$. This amount would nominally suggest
recent formation, if one adopts only passive evolution after a single
burst of star formation, but oddly, the colors of these galaxies were
moderately red. This result was confirmed by a fourth\cite{Koo02},
purely photometric study of the luminosities and colors of luminous
bulge subcomponents rather than the entire galaxy.  While the
size-luminosity relation indicated nearly the same amount of
luminosity evolution ($\sim 1.5$mag), virtually all bulges were
nevertheless found to be very red (restframe $U-B \sim 0.5$). A
possible explanation for this paradox would be that a very metal-rich,
old stellar population is later contaminated by small amounts of bluer
young or metal-poor stars.

\subsection{Luminous Blue Compact Galaxies}

Though spatially-resolved rotation curves are preferred, most faint
galaxies are too small to yield more than linewidths as kinematic
data.  Except for galaxies bright enough to yield {\it absorption}
linewidths, emission lines are used (see contribution by B. Weiner).
Assuming linewidths are reliable measures of the true gravitational
potential (after an upward correction of 40\%
\cite{Rix97},\cite{Tell93} to match HI or H$\alpha$ values of
kinematics), and adding HST sizes, we are able to obtain masses.

\begin{figure}[b]
\begin{center}
\includegraphics*[bb = 100 450 500 700,height=2.5in]{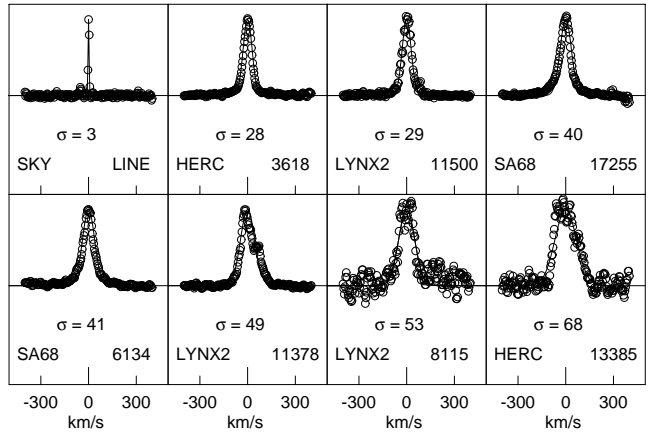}
\end{center}
\caption[]{Panel of HIRES \cite{Vogt94} emission line profiles from a sample of
luminous ($L^*$) blue compact galaxies at redshifts
$z\sim$0.1--0.7 \cite{Guz96}. 
The $\sigma$ are the FWHM/2.35 velocity widths in km s$^{-1}$. The dynamic
range of the M/L in rest-frame B for these compact galaxies spans a
factor of 45. 
The HIRES instrumental profile is shown in the upper-left panel.
}
\label{fig2}
\end{figure}

The masses of blue compact galaxies \index{BCG}\index{blue compact galaxies} 
have been found to be especially
interesting.  For some, we find that luminosity can be a poor gauge of
their masses. Though many have the luminosities of massive
galaxies(L$^*$), their velocity widths ($\sigma$) may be smaller than
30 km-s$^{-1}$ as seen in the line profiles shown in Fig.\ref{fig2}
\cite{Koo95b}, which were measured using the high resolution echelle
spectrograph (HIRES: \cite{Vogt94}) . The resultant masses are very
small and yield M/L ratios that span a wide range \cite{Guz96},
\cite{Guz97}, \cite{Phi97}.  Our results suggest that some luminous
blue compacts may be the progenitors of quiescent spheroidals today;
that the down-sizing scenario \cite{Cow96} may apply to these galaxies
\cite{Guz97}; and that such galaxies seen at $z < 1$ may be
lower-redshift counterparts to the Lyman-drop galaxies seen at higher
redshifts $z \sim 3$ \cite{Stei96a}.  A key point is that optical
luminosities and mass are seen to be poorly correlated for this
sample, i.e. stable and constant M/L may be a poor assumption at least
for some classes of galaxies. This result clearly demonstrates the
necessity, usefulness, and promise of kinematics as an important new
dimension to discern the evolution of different galaxy populations.

\subsection{High Redshift $z \sim 3$ Galaxies}

A major advance with Keck has been the dramatic demonstration that
galaxies chosen by multicolor photometry to be at very high redshift
($z \sim 3$) are confirmed to be so spectroscopically
\cite{Stei96a}. The DEEP team has extended the pioneering efforts in
the Hubble Deep Field (HDF) \cite{Stei96b} by pushing over one
magnitude fainter; using redder ``dropouts'' to reach higher redshifts
and higher levels of completeness; and adopting higher spectral
resolutions to improve kinematic measurements \cite{Low97}.  
Small motions observed from spatially-unresolved 
velocity widths \cite{Low97} \cite{Pet01}
and spatially-resolved kinematics indicate that some  $z \sim 3$
galaxies appear to be small-mass systems that become dwarfs today or that
later merge to form more massive galaxies \cite{Low97}, instead of
being only the cores of massive galaxies in formation (which should yield
high motions).

\section{Summary} 

The main theme that arises from our DEEP pilot programs is that galaxy
evolution is a complex problem.  Galaxies are diverse in size,
luminosity, structure, etc.; are composed of subcomponents which may
experience different star formation and dynamical histories and
evolution; and reside in a wide range of environments involving
different physical mechanisms for their evolution. We have
established that
kinematics are both feasible with 8-10~m class telescopes and valuable
for understanding distant galaxies. For example, we find relatively
little evolution in the Tully-Fisher relation 
or disk surface brightness \cite{Sim99} to redshifts $z \sim 1$, as
well as little evidence for evolution in the fundamental plane, volume
density, or luminosity beyond that expected from passive evolution for
early-type galaxies to $z \sim 1$. The colors of the spheroids and
bulges are, however, redder than expected and thus a puzzle. On the
other hand, luminous blue compact galaxies appear, whether at low
redshifts $z < 1$ or at high redshifts $z \sim 3$, to have very low
dynamical masses and are suggested to be possible progenitors of
quiescent low-mass spheroidals today or the building blocks of larger
galaxies rather than massive ellipticals undergoing formation via
monolithic collapse.

The lessons from our DEEP phase-one pilot programs indicate great
promise for our main survey DEEP2 of 50,000 galaxies. Such large
numbers are vital for analysis after subdivision of the full sample by
a wide range in luminosity, size, M/L, structure, redshift, and
environment.  More relevant to this conference, we are optimistic that
the kinematic data will yield new studies that rely on mass, including
mass functions; M/L functions; Tully-Fisher, Fundamental Plane, and
other scaling law evolution; mergers rates; dark matter distributions
(halo vs disk; large scale structure vs mass); and precision cosmology
(velocity function vs volume tests; estimates of the equation of state
\cite{New00}).

{\bf Acknowledgements:} DEEP was initiated by the Berkeley Center for
Particle Astrophysics (CfPA), and has been supported by various other
NSF, NASA, UC, and STScI grants. The senior members of DEEP have
managed the project, but I would like to give special thanks to our
talented pool of more junior astronomers over the years (see DEEP URL
for names), without whom the results presented here would not have
been possible.

%

\end{document}